\documentclass[letter,bibyear]{aa}
\usepackage[varg]{txfonts}
\usepackage{graphicx}
\def\S{\mbox{S255IR~NIRS\,3}}

\def\Msun{\mbox{$M_\odot$}}

\def\kms{\mbox{km~s$^{-1}$}}

\def\Tb{T_{\rm B}}
\def\Tbt{\tilde{T}_{\rm B}}

\def\tho{\theta_0}

\def\ro{r_0}
\def\yo{y_0}
\def\ym{y_{\rm m}}

\def\rmax{r_{\rm m}}

\def\vo{\varv_0}

\def\no{n_0}
\def\To{T_0}
\def\xo{x_0}
\def\tauo{\tau_0}

\begin{document}

\title{
Radio outburst from a massive (proto)star
}
\subtitle{III. Unveiling the bipolarity of the radio jet from \S
\thanks{Based on observations carried out with the VLA.}}
\author{
        R. Cesaroni\inst{1}
        \and
        L. Moscadelli\inst{1}
        \and
        A. Caratti o Garatti\inst{2,3}
        \and
        J. Eisl\"offel\inst{4}
        \and
        R. Fedriani\inst{5}
        \and
        R. Neri\inst{6}
        \and
        T. Ray\inst{3}
        \and
        A. Sanna\inst{7}
        \and
        B.~Stecklum\inst{4}
}
\institute{
 INAF, Osservatorio Astrofisico di Arcetri, Largo E. Fermi 5, I-50125 Firenze, Italy
           \email{riccardo.cesaroni@inaf.it}
\and
 INAF, Osservatorio Astronomico di Capodimonte, via Moiariello 16, I-80131 Napoli, Italy
\and
 Dublin Institute for Advanced Studies, School of Cosmic Physics, Astronomy \& Astrophysics Section, 31 Fitzwilliam Place, Dublin 2, Ireland
\and
 Th\"uringer Landessternwarte Tautenburg, Sternwarte 5, D-07778 Tautenburg, Germany
\and
 Instituto de Astrof\'{\i}sica de Andaluc\'{\i}a, CSIC, Glorieta de la Astronom\'{\i}a s/n, E-18008 Granada, Spain
\and
 Institut de Radioastronomie Millim\'etrique (IRAM), 300 rue de la Piscine, F-38406 Saint Martin d’H\`eres, France
\and
 INAF, Osservatorio Astronomico di Cagliari, Via della Scienza 5, I-09047 Selargius (CA), Italy
}
\offprints{R. Cesaroni, \email{riccardo.cesaroni@inaf.it}}
\date{Received date; accepted date}

\abstract{
We report new Very Large Array high-resolution observations of the
radio jet from the outbursting high-mass star \S. The images at 6, 10,
and 22.2 GHz confirm the existence of a new lobe emerging to the SW and
expanding at a mean speed of $\sim$285~\kms, about half as fast as
the NE lobe. The new data allow us to reproduce both the morphology and the
continuum spectrum of the two lobes with the model already adopted in our previous
studies. We conclude that in all likelihood both lobes are powered by
the same accretion outburst. We also find that the jet is currently fading
down, recollimating, and recombining.
}
\keywords{Stars: individual: \S -- Stars: early-type -- Stars: formation -- ISM: jets and outflows}

\maketitle

\section{Introduction}
\label{sint}

Circumstellar discs and jets are believed to play a crucial role in the
formation of stars of all masses. While many objects of
this type are known  for solar-type stars, disc--jet systems associated with young early-type stars
have only been revealed and studied in detail for a limited number of cases
(Beltr\'an \& de Wit~\cite{beldew}).
The advent of the Atacama Large Millimeter and submillimeter Array (ALMA)
as well as the upgrade of the \textit{Karl Jansky} Very Large Array (VLA)
have contributed to improving our knowledge in this field.
As a consequence of such a substantial technical improvement, it is
not only possible to perform large surveys of many disc--jet candidates,
but also to monitor their emission and reveal changes in both the flux
density and morphology due to the expansion and evolution of the jet.

With all of the above in mind, we recently conducted an observational
campaign in a variety of tracers (Stecklum~\cite{steck16,steck21};
Caratti o Garatti et al.~\cite{cagana}; Moscadelli et al.~\cite{mosca17};
Hirota et al.~\cite{hiro21}) of the young stellar object (YSO) \S\
(distance 1.78~kpc; Burns et al.~\cite{burns16}), which is one of the
few cases of an accretion outburst detected towards a young massive star
(Tapia et al.~\cite{tapia}; Hunter et al.~\cite{hunt17,hunt21}; Burns et
al.~\cite{burns20,burns23}; Chen et al.~\cite{chen21}). As part of this campaign,
we made multi-epoch observations of the source at centimetre
(with the VLA) and millimetre (with the NOEMA and ALMA interferometers)
wavelengths (Cesaroni et al.~\cite{cesa18}, hereafter Paper~I; Cesaroni et
al.~\cite{cesa23}, hereafter Paper~II), showing that the emission
from the ionised jet also underwent a burst, although with a delay of about
1 year with respect to the infrared burst detected by Caratti o Garatti et
al.~(\cite{cagana}). The flux-density variation is explained with a
model that takes into account the expansion of a conical ionised jet. This model
can reproduce both the continuum spectrum and the morphology of the jet.

One of the results obtained in our studies is that although the jet 
initially consists of only the NE lobe (see also Fedriani et al.~\cite{fedr23}), a SW lobe appears between approximately 22 and 35~months after the
onset of the radio outburst, in July~2016. We propose that
this new lobe is powered by the same accretion outburst (which began in mid-June
2015; see Caratti o Garatti et al.~\cite{cagana}) as the NE lobe, despite
the significant time lag between the two. In Paper~II, we speculated that
the delay is due to an inhomogeneous distribution of the material around
the YSO, with the gas density to the SW being greater than that to the NE
of the star. To shed light on these issues, we performed new VLA subarcsecond-resolution observations of \S\ to investigate the outbreak of the SW lobe,
as well as its intensity variation with time. In the following, we report the results obtained.

\section{Observations}
\label{sobs}

We used the A-array configuration of the VLA on July 21, 2023 (project
code: 23A-021), to image the continuum emission of the radio jet from \S\
at three frequencies: 6, 10, and 22.2~GHz, which correspond
to the C, X, and K bands, respectively.

The Wideband Interferometric Digital Architecture (WIDAR) correlator was used
in dual polarisation mode. The total observing bandwidth (per polarisation)
was 4~GHz in the C and X bands and 8~GHz in the K band. The primary flux
calibrator was 3C48 and the phase calibrators were J0559$+$2353 in the C
and X bands, and J0539+1433 in the K band.

We made use of the calibrated data set provided by the NRAO pipeline
and subsequent inspection of the data and imaging were performed with
the CASA\footnote{The Common Astronomy Software Applications software
can be downloaded at http://casa.nrao.edu} package, version 5.6.2-2.
For the continuum images, we adopted natural weighting in the X and
K bands and `Briggs' weighting with `robust=0.5' in the C band, which
provides a good compromise between angular resolution and sensitivity
to extended structures. The half-power widths and position angles
of the synthesised beams are 0\farcs30$\times$0\farcs28, 0\fdg3
at 6~GHz, 0\farcs23$\times$0\farcs22, -1\fdg3 at 10~GHz, and
0\farcs10$\times$0\farcs093, 0\degr\ at 22.2~GHz. The typical noise is
$\sim$0.02~mJy/beam in all bands for an on-source time of 26~min in the
C and X bands and 20~min in the K band. The calibration uncertainty is
estimated to be 15\% in all bands.

\section{Results and analysis}
\label{sres}

\begin{figure}
\centering
\resizebox{7.0cm}{!}{\includegraphics[angle=0]{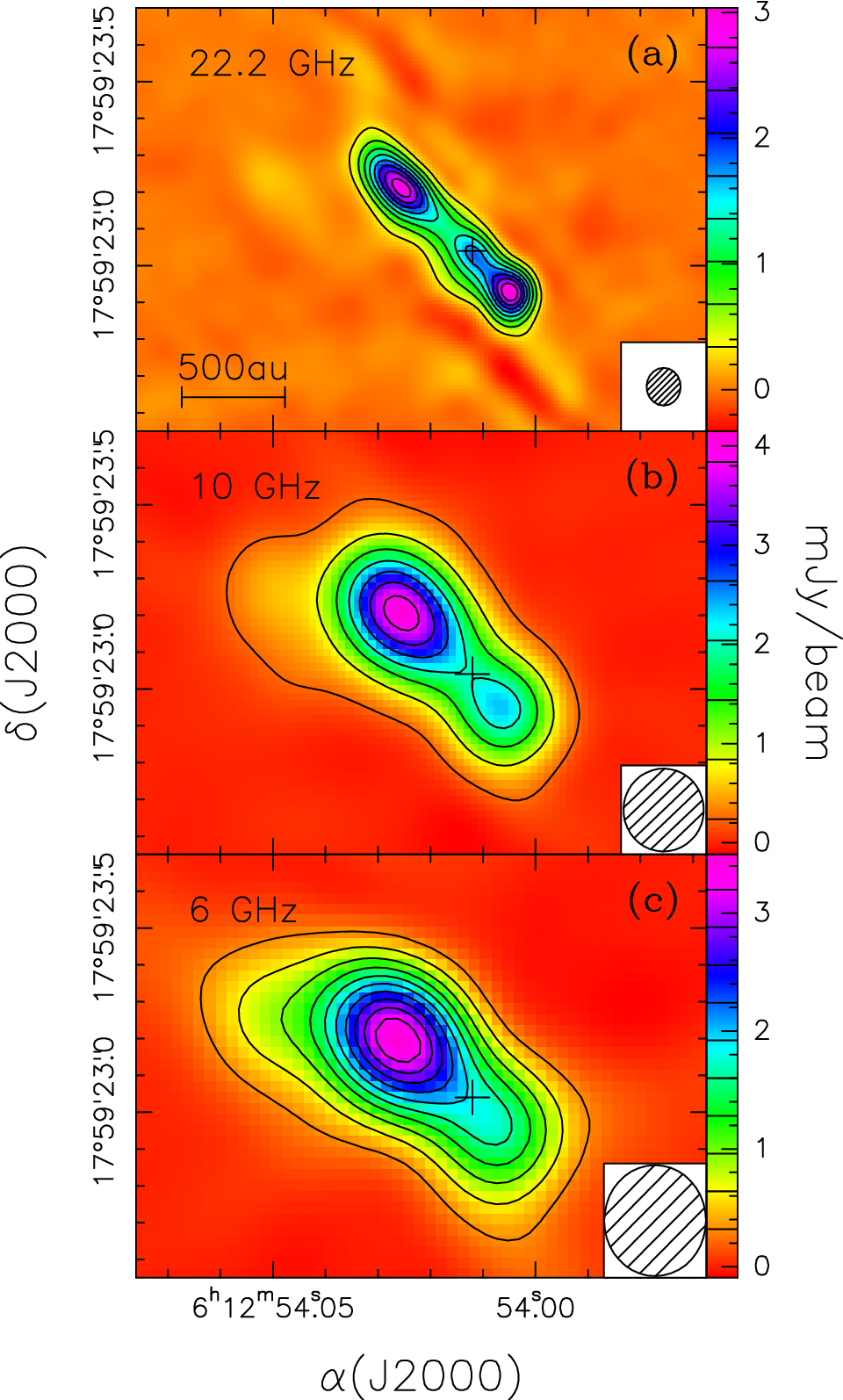}}
\caption{
Maps of the continuum emission from \S\ obtained with the VLA.
{\bf a.} Data acquired at 22.2~GHz on July 21, 2023. Contour levels are
quantified by the colour scale to the right of the panel. The ellipse
in the bottom right is the half-power width of the synthesised beam.
The cross marks the peak of the 3~mm continuum emission from Paper~II,
which we assume to be the position of the star.
{\bf b.} Same as top panel, but for the 10~GHz emission.
{\bf c.} Same as top panel, but for the 6~GHz emission.
}
\label{fmaps}
\end{figure}

\begin{figure}
\centering
\resizebox{8.0cm}{!}{\includegraphics[angle=0]{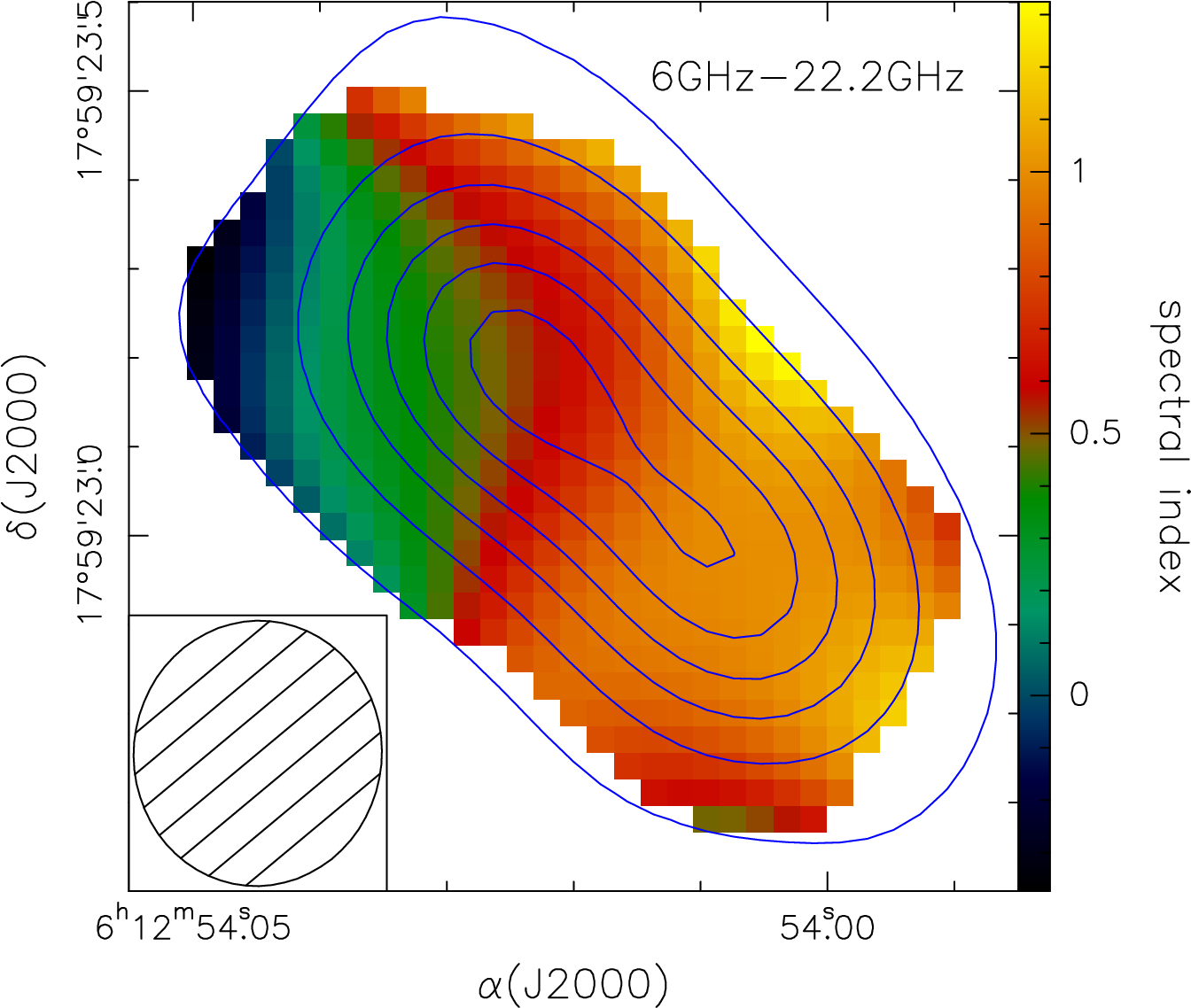}}
\caption{
Map of the spectral index obtained from the ratio between the 6~GHz and
22.2~GHz images. The contours are a map of the 22.2~GHz continuum emission
obtained with the same clean beam as the 6~GHz image. Contour levels range
from 0.5 to 5.5 in steps of 1~mJy/beam.  The ellipse in the bottom right
is the half-power width of the synthesised beam.
}
\label{fspi}
\end{figure}

Figure~\ref{fmaps} shows the maps obtained at the three frequencies.
It is quite evident that the jet structure is bipolar with a prominent lobe
to the NE. The ratio between the intensities of the two lobes appears to
change with frequency, which suggests different opacities, and therefore column
densities, of the two lobes. To better investigate this aspect, we plot in
Fig.~\ref{fspi} the maps of the spectral index computed from the ratio between
the 6~GHz and 22~GHz maps after cleaning the latter with the same beam as
the former. These maps, obtained from consecutive observations, are only
compared in the higher intensity contours to minimise possible systematic uncertainties
(see Sanna et al.~\cite{sanna18}). One sees that the spectral index is
steeper to the SW than to the NE, which is consistent with the free-free emission
being thicker. This result agrees with the idea proposed in Paper~II that
the expansion of the SW lobe could be curbed by a density enhancement in
that direction.

\subsection{Expansion of the jet lobes}

While in Paper~II we established that the NE lobe is expanding
with velocity decreasing with time, the data obtained in that study were
insufficient to gain insight into the expansion of the SW lobe. With the
new VLA image at 22.2~GHz, it is now possible to shed light
on this aspect in particular. For this purpose, it is necessary to compare the ALMA maps
acquired at 3~mm in 2019 and 2021 with the VLA 1.3~cm map obtained in 2023.

A caveat with this approach is that the 3~mm continuum is heavily
contaminated by emission from dust grains, especially in the SW lobe, whereas
the 1.3~cm continuum is mostly due to pure free-free emission. To
circumvent this problem, we applied Eq.~(3) of Paper~II, which
allows us to compute the expected fraction of the flux density contributed
by free-free emission based on the measured spectral index at 3~mm. In
practice, we corrected the flux in each pixel of the 3~mm continuum maps
(see Fig.~1 of Paper~II) using the corresponding spectral index (see Fig.~7
of Paper~II). In the calculation, we adopted the dust spectral index of 2.08
derived in Eq.~(4) of Paper~I. We note that for some pixels, especially
in the NE lobe, the spectral index lies below the minimum value of --0.1
expected for free-free or dust emission. In these cases, we used --0.1 for
the calculation.

\begin{figure}
\centering
\resizebox{8.0cm}{!}{\includegraphics[angle=0]{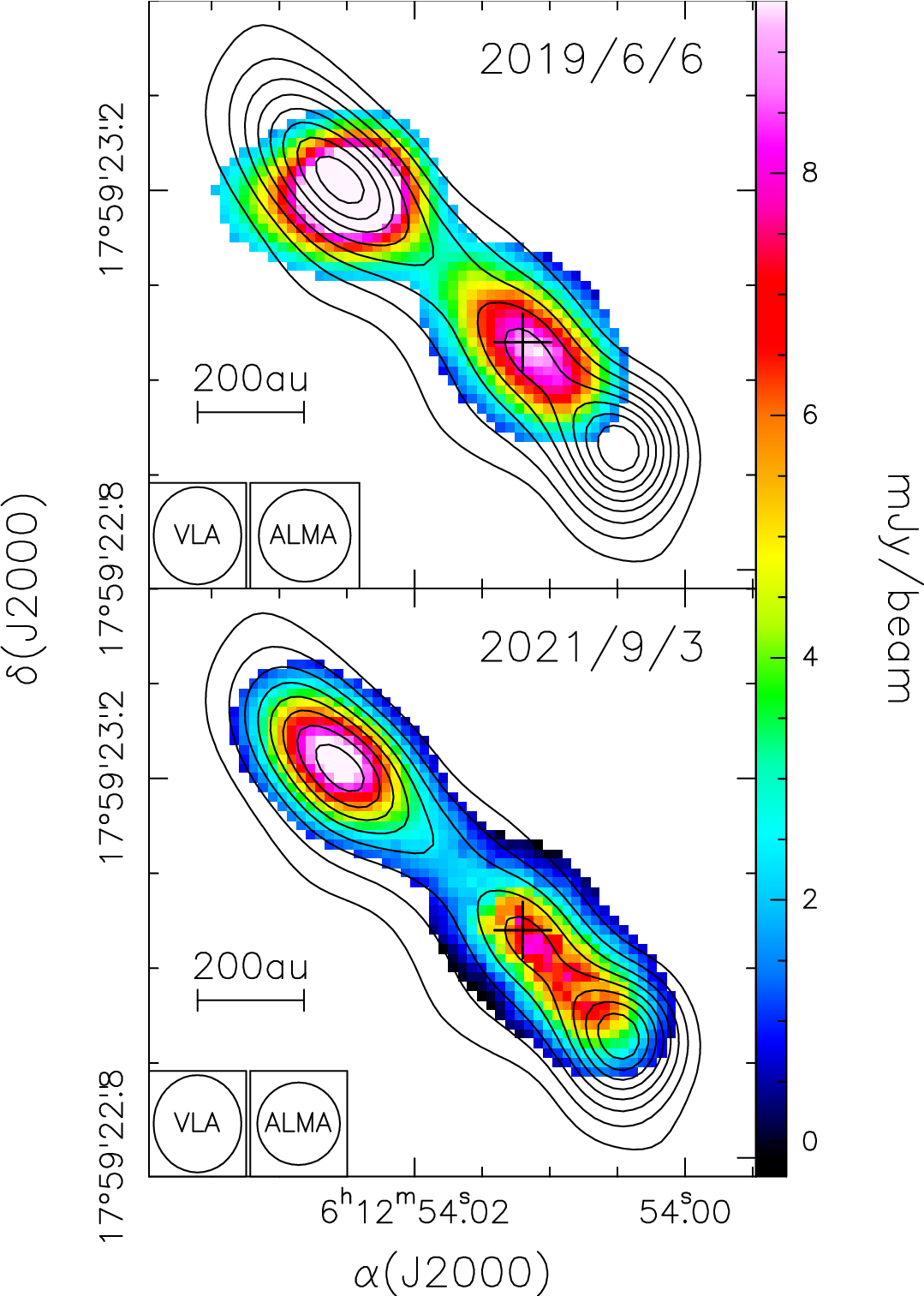}}
\caption{
ALMA maps of the 92.2~GHz emission (colour scale, saturated to emphasise the emission
from the SW lobe), after subtracting the dust contribution (see text).
The contours are the map of the 22.2~GHz continuum emission obtained in
2023. Contour levels range from 0.34 to 2.72 in steps of 0.34~mJy/beam.
The ellipses in the bottom left are the half-power widths of the synthesised beams.
{\bf a.} Obtained from the ALMA data acquired on June 6, 2019 (see Paper~II).
{\bf b.} Same as top panel, but for the data obtained on September 3, 2021.
}
\label{ffree}
\end{figure}

The `pure' free-free maps obtained with this method are shown in
Fig.~\ref{ffree}, where a comparison is also made with our new 1.3~cm image.
Although it is not possible to identify a precise peak of emission in
the 2021 map, there is no doubt that the SW lobe becomes more elongated
with time. A mean projected expansion speed of $\sim$285~\kms\ is estimated from the
separation of $\sim$0\farcs14=248~au between the SW peak of the 2019 image
and that of the 1.3~cm map, and the corresponding time interval of 1506~days.
We note that this value is close to the true expansion speed because the
inclination of the jet on the plane of the sky is $\sim$10\degr\ (see Paper~II).
Such a speed is about half that estimated for the NE peak in
Paper~II, of namely $\sim$440~\kms, which suggests that the expansion
has indeed been slowed down more on the SW side than on the NE side. This evidence
supports the idea that the SW lobe was also powered by the same outburst
as the NE lobe, but was impeded at the beginning of its expansion.

\subsection{Evolution of the jet emission}

\begin{table}
\caption[]{
Flux densities of the free-free continuum emission from the jet lobes measured in 2019, 2021, and 2023.
}
\label{tflux}
\begin{tabular}{c|cccccccc}
\hline
\hline
year && \multicolumn{3}{c}{2023} && 2019 && 2021 \\
 \cline{3-5} \cline{7-7} \cline{9-9}
$\nu$ (GHz) && 6 & 10 & 22.2 && 92.2 && 92.2 \\
\hline
NE (mJy)  && 4.8 & 6.9 & 7.8 && 22.8 && 18.8 \\
SW (mJy)  && 2.4 & 3.3 & 6.7 && 13.5$^{(a)}$ && 14.6$^{(b)}$ \\
\hline
\end{tabular}

\vspace*{1mm}
$^{(a)}$ Obtained after subtracting the dust emission from the total flux of 23.2~mJy. ~
$^{(b)}$ Obtained after subtracting the dust emission from the total flux of 23.5~mJy.

\end{table}

\begin{figure}
\centering
\resizebox{8.5cm}{!}{\includegraphics[angle=0]{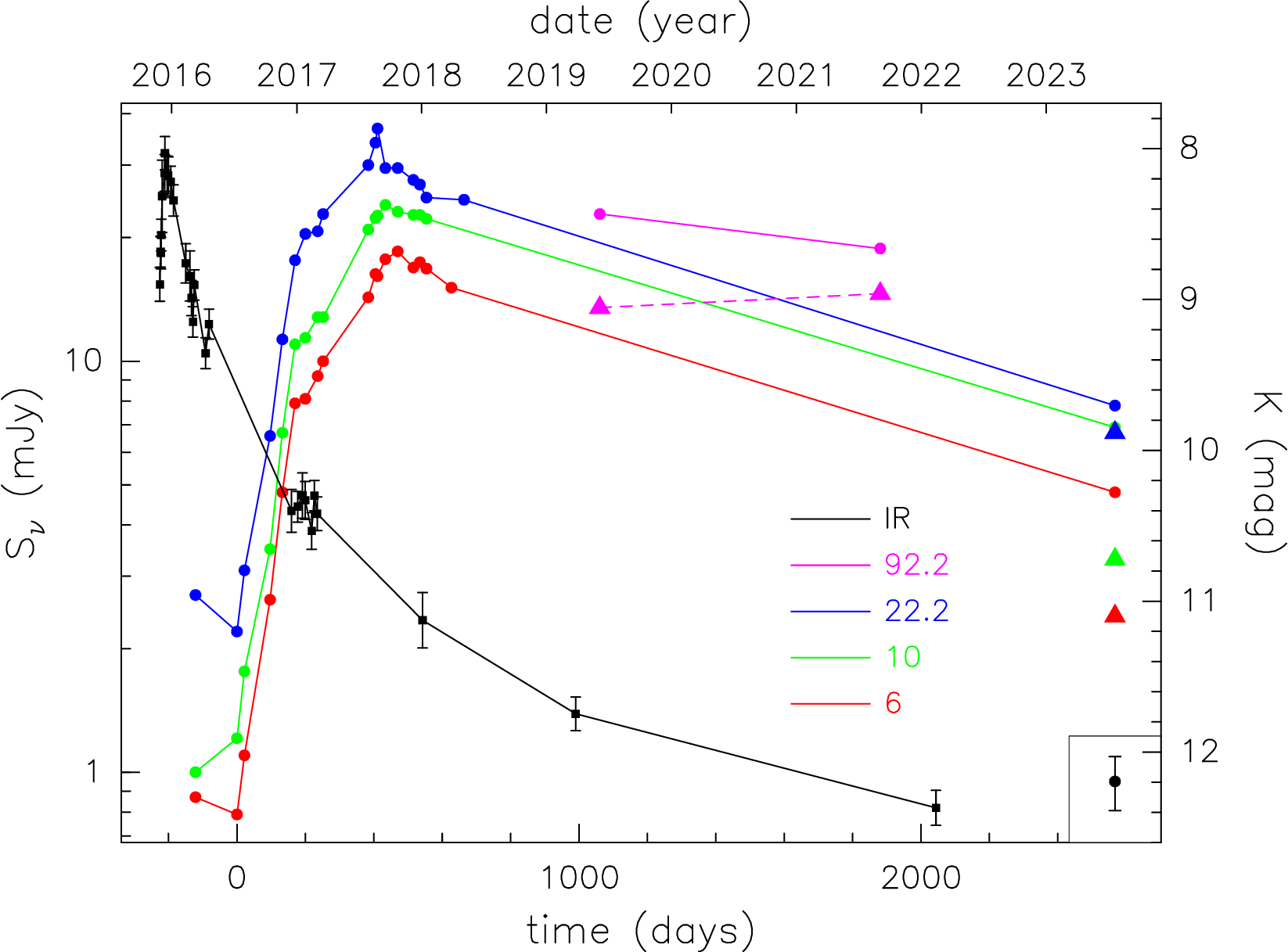}}
\caption{
Plot of the free-free continuum emission from the radio jet at 6, 10,
and 22.2~GHz (coloured curves) as a function of time measured from the
onset of the radio outburst on July 10, 2016 (see Paper~I). Circles and
solid lines indicate the emission from the NE lobe, while triangles and
the dashed line denote the emission from the SW lobe. The error bar in
the bottom right corresponds to an uncertainty of 15\%.
The black curve with error bars is the light
curve of the accretion outburst obtained at K band in the near-infrared
(see Fedriani et al.~\cite{fedr23}).
}
\label{ffvst}
\end{figure}

In Table~\ref{tflux}, we give the flux densities of the free-free continuum
emission from each jet lobe obtained from the new VLA data. We also list the
`pure' free-free fluxes at 92.2~GHz computed as previously explained from
the ALMA data presented in Paper~II. We used the flux densities in this table as
well as those already obtained in Paper~II to produce the
light curves of both lobes, which are shown in Fig.~\ref{ffvst}.
At all frequencies, we see that the free-free emission from the NE lobe
(circles connected by solid lines) has decreased by a factor of about 3
at the last epoch with respect to the previous measurement. For the SW
lobe (triangles and dashed line), we have single-epoch data at centimetre
wavelengths, but the previous two measurements at 3~mm indicate a slight
increase in the flux density. To compare the emission from the SW lobe in
2021 with that in 2023 is not trivial because the two measurements available
are taken at different frequencies. However, it is very reasonable to assume
that the 92.2~GHz flux is close to that measured at 22.2~GHz, because
at the last epochs of the monitoring in Paper~II (Fig.~4) the spectrum appears to be
approximately flat above 22.2~GHz and the turn-off frequency $\nu_{\rm m}$ (see
Fig.~A.1 of Paper~I) appears to decrease with time. It is therefore likely that
the flux from the SW lobe has dropped by a factor of about 2 after year 2021.

We conclude that the jet activity has been weakening since 2021, as
expected given that the energy input provided by the accretion event has come to
an end. This can be appreciated in Fig.~\ref{ffvst} by comparing the radio
light curves with the IR light curve of the accretion outburst obtained from
Table~B.1 of Fedriani et al.~(\cite{fedr23}). From this comparison, the
time lag of $\sim$600~days between the peaks of the two curves
clearly stands out, and is roughly consistent with the delay of $\sim$400~days (see Paper~I) between
the onset of the NIR burst and that of the radio burst. Also, both curves
are characterised by a rapid increase and a slow decline, but the timescales of the IR light curve are shorter, as expected for radiative processes
(the IR burst) as opposed to hydrodynamical processes (the jet expansion).

\subsection{Modelling the spatial and spectral emission from the jet}

With the new VLA data, for the first time since the onset of the radio
outburst in \S, we can both resolve the jet structure and measure the
continuum flux at three frequencies. We can therefore fit both the jet morphology
and the spectrum at the same time. For this purpose, we adopted the
same model described in Paper~II, namely a slightly modified version of the
Reynolds~(\cite{reyn}) model. In summary, this consists of a conical jet
with opening angle $\tho$ and vertex coinciding with the position of the
star. The ionised gas temperature, $\To$, ionisation fraction, $\xo$, and
internal velocity, $\vo$, do not depend on the distance from the star,
$r$, whose projection on the plane of the sky is $y=r\,\cos\psi$, with
$\psi=10\degr$ being the inclination of the jet on the plane of the sky (see
Paper~II). The gas number density is $n=\no(\ro/r)^2$. The ionised gas is
confined between an inner radius $\ro$ and an outer radius $\rmax$. All
parameters but $\To$ may change with time.

The model fit to the 22.2~GHz map of the jet was performed assuming that the whole jet
is optically thin. This implies that the brightness temperature is
$\Tb=\To\tau(y)=\To\tauo(\ro/y)^3$ (see Eq.~B.3 of Paper~II)
and one can fit the map normalised with respect to the peak
intensity $\Tbt\equiv\Tb/\Tb(\yo)=(\yo/y)^3$. In this way, only three
free parameters are left, that is, $\tho$, $\yo$, and $\ym$. The best fit is
obtained, after convolving the model to the synthesised beam of the
observations, by minimising the expression
$\sum_i \left(\tilde{T}_{\rm B}^i({\rm model})-\tilde{T}_{\rm B}^i({\rm data})\right)^2$,
where $i$ is a generic pixel of the image. We then fix $\tho$, $\yo$,
and $\ym$ to the best-fit values and vary $\tauo$ to minimise Eq.~(6)
of Paper~II.

The fit to the spectrum depends on four parameters, $\tho$, $\yo$, $\ym$, and
$\Lambda$, where the latter is defined as $\Lambda=\xo\,\dot{M}/\vo$ with
$\dot{M}$ being the mass-loss rate of the jet. It is worth noting that $\Lambda$
can be expressed as a function of $\tauo$ through Eqs.~(B.6) and~(B.11)
of Paper~II. The best fit is obtained by minimising Eq.~(10) of Paper~I.
As the two lobes evolve in a different way, it is convenient to discuss
the fit to each of them separately.

\subsubsection{NE lobe}

\begin{figure}
\centering
\resizebox{8.8cm}{!}{\includegraphics[angle=0]{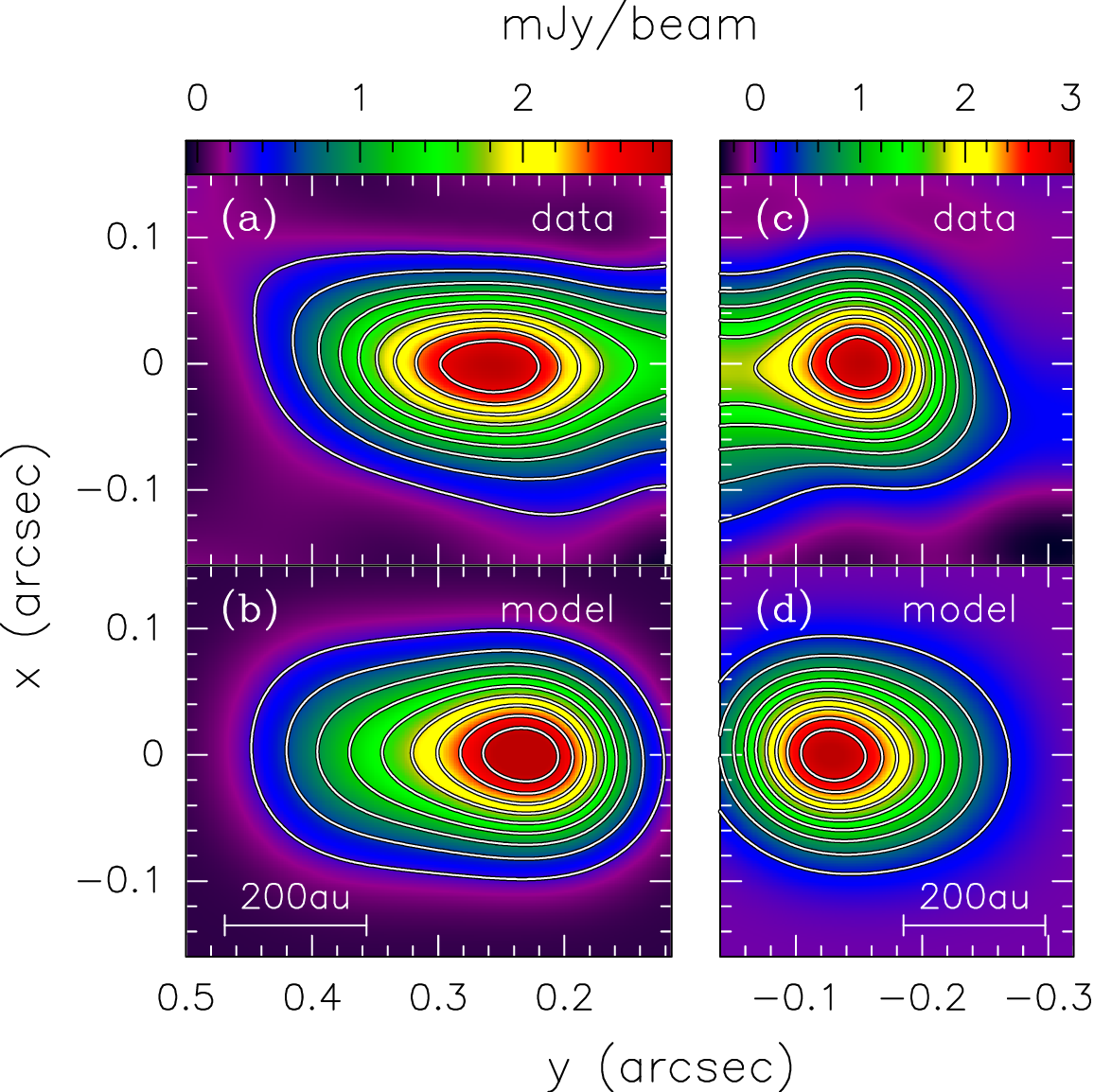}}
\caption{
Comparison between observed and model maps of the 22.2~GHz continuum emission
from the jet lobes.
{\bf a.} Map of the NE lobe of the jet,
rotated by 42\degr\ counterclockwise. Contour levels range from 10\% to
90\% in steps of 10\% of the peak emission.
{\bf b.} Best-fit model of the NE lobe.
Contour levels are the same as for the data.
{\bf c.} Same as panel `a', but for the SW lobe, rotated by 48\degr\ counterclockwise.
{\bf d.} Same as panel `b', but for the SW lobe.
}
\label{fmorf}
\end{figure}

As a first step, we fitted the morphology of the NE lobe as explained above.
Figure~\ref{fmorf}a shows the 22.2~GHz map of the NE lobe rotated by
42\degr\ counterclockwise for the sake of comparison with the model fit
in Fig.~\ref{fmorf}b. The corresponding best-fit parameters are given in
Table~\ref{tfits}.

\begin{table*}
\caption[]{
Parameters of the best fits to the maps and spectra of the jet lobes.
}
\label{tfits}
\begin{tabular}{c|cccccccccc}
\hline
\hline
fitted   && \multicolumn{4}{c}{NE lobe} && \multicolumn{4}{c}{SW lobe} \\
\cline{3-6} \cline{8-11}
data     && $\tho$ & $\yo$ & $\ym$ & $\Lambda$ && $\tho$ & $\yo$ & $\ym$ & $\Lambda$ \\
         && (deg) & (au) & (au) & (\Msun~yr$^{-1}$/(\kms)) && (deg) & (au) & (au) & (\Msun~yr$^{-1}$/(\kms)) \\
\hline
map      && $7.9^{+1.0}_{-1.2}$ & $329^{+5}_{-6}$ & $760^{+20}_{-20}$ & $(6.22^{+0.03}_{-0.03})\times10^{-9}$ &
         & $11^{+2.5}_{-3.4}$  & $164^{+5}_{-6}$ & $424^{+22}_{-14}$ & $(4.95^{+0.03}_{-0.03})\times10^{-9}$ \\
spectrum && $9.1^{+5.6}_{-2.4}$ & 329$^{(a)}$       & 715$^{(b)}$   & $(7.9^{+4.2}_{-1.5})\times10^{-9}$    &
         & $9.9^{+2.9}_{-1.9}$ & 164$^{(a)}$       & 424$^{(a)}$     & $(5.1^{+1.4}_{-1.0})\times10^{-9}$    \\
\hline
\end{tabular}

\vspace*{1mm}
$^{(a)}$ Assumed equal to the value obtained from the fit to the map. ~
$^{(b)}$ Computed from Eq.~(7) of Paper~II.

\end{table*}

\begin{figure}
\centering
\resizebox{7.0cm}{!}{\includegraphics[angle=0]{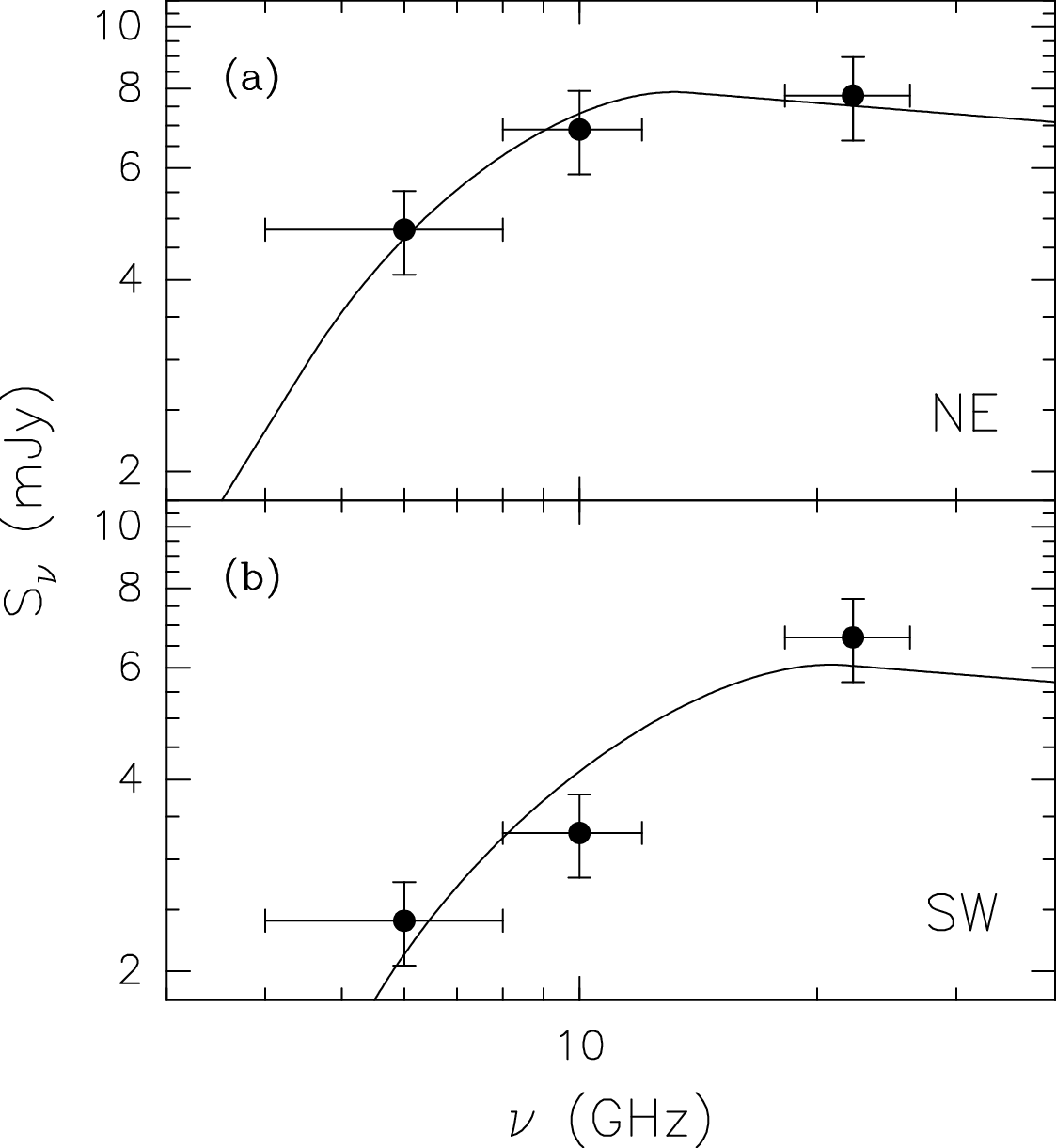}}
\caption{
Spectra of the continuum emission from the jet lobes.
The vertical bars correspond to a 15\% error on the flux density, whereas
the horizontal bars indicate the bandwidth covered at each frequency.
The curves are the best fits obtained with the model described in the text.
{\bf a.} NE lobe.
{\bf b.} SW lobe.
}
\label{fspec}
\end{figure}

The second step is the fit to the spectrum. As only three data points
are available, we prefer to reduce the number of free parameters by fixing
$\yo$ to the value obtained from the fit to the map, 329~au, while $\ym$
is computed from Eq.~(7) of Paper~II. The best fit is obtained for
the values of $\tho$ and $\Lambda$ in Table~\ref{tfits} and is shown in
Fig.~\ref{fspec}a. We note that the values of $\tho$ and $\Lambda$ are in
good agreement with those derived from the fit to the map, which lends
support to the results obtained.
We stress that assuming $\ym=760$~au, as obtained from the fit to the map,
would change $\tho$ and $\Lambda$ by only $\sim$10\%.

\begin{figure}
\centering
\resizebox{7.0cm}{!}{\includegraphics[angle=0]{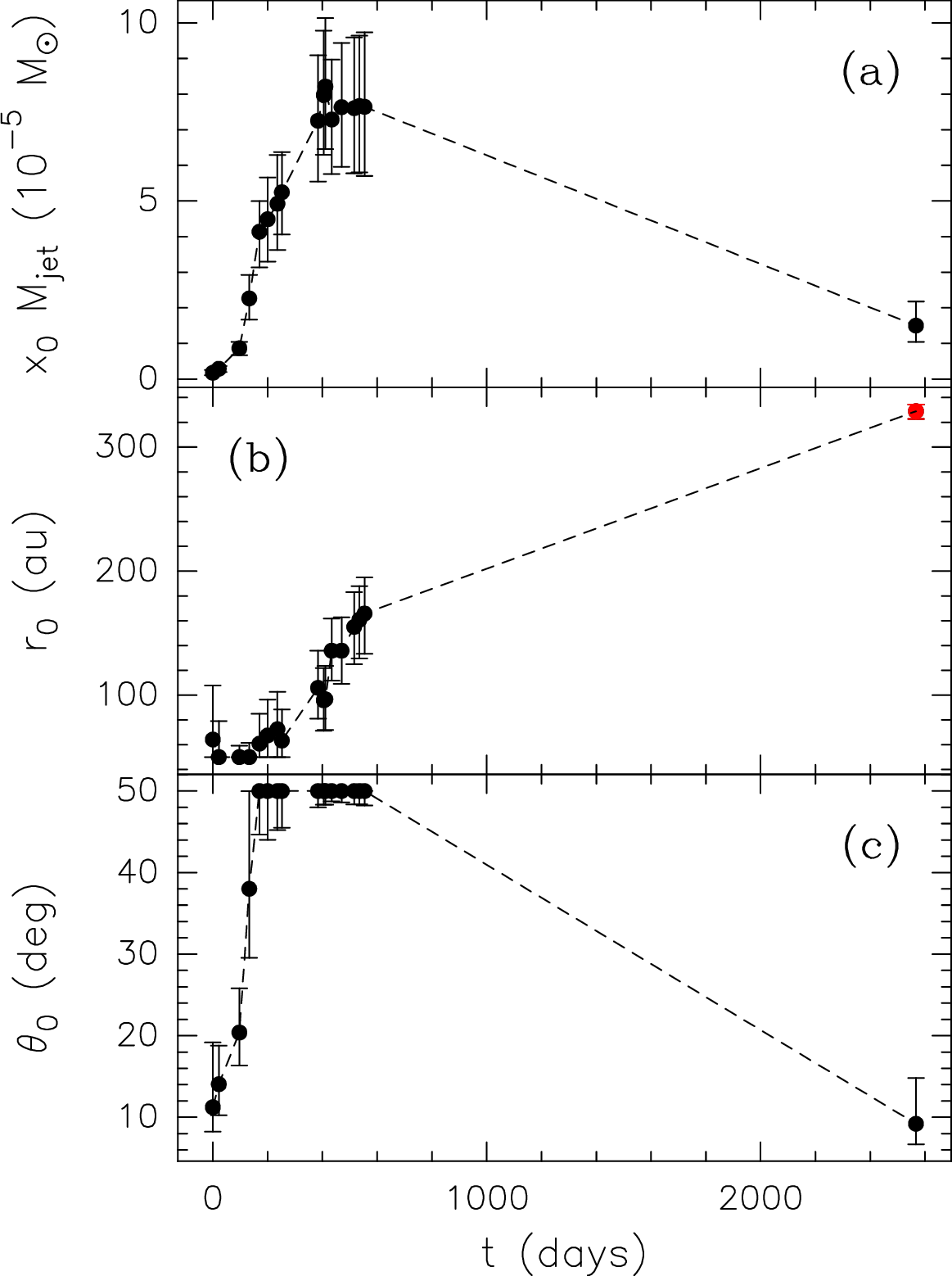}}
\caption{
Parameters obtained from the best fit to the continuum spectrum of the NE lobe
as a function of time.
{\bf a.} Ionised mass of the jet.
{\bf b.} Inner radius of the NE lobe. The red point was obtained by
fitting the map of the 22.2~GHz emission (Fig.~\ref{fmorf}b).
{\bf c.} Opening angle of the jet.
}
\label{fpars}
\end{figure}

Figure~\ref{fpars} shows the best-fit values of $\tho$ and $\ro$, and the
ionised jet mass computed from Eq.~(B.13) of Paper~II as a
function of time for all the data already presented in Paper~II plus the
new data analysed in the present study. The conclusion is that the inner
radius keeps expanding, while the jet recollimates, as indicated by the
dramatic decrease in $\tho$. Furthermore, the ionised gas recombines,
because the decrease in $\xo M_{\rm jet}$ cannot be due to a decrease in
$M_{\rm jet}$, which remains constant after the inner radius begins to
grow and the jet feeding stops (see also Paper~II).

\subsubsection{SW lobe}

The model fit to the map and spectrum of the SW lobe was performed in the
same way as for the NE lobe, with the only difference being that we do not have
an expression analogous to Eq.~(7) of Paper~II to calculate $\ym$ as a
function of time. For this reason, when fitting the spectrum, we assumed
$\ym$ equal to the value obtained from the fit to the 22.2~GHz map.
The best-fit parameters are given in Table~\ref{tfits}. Also in this case,
as for the NE lobe, there is good agreement between the values of $\tho$
and $\Lambda$ estimated from the map and those derived from the spectrum.

While the opening angle and $\Lambda$ are similar to those of the NE lobe,
the inner and outer radii are significantly smaller, a result consistent
with the SW lobe expansion being hindered by the environment. The ionised
jet mass is $\sim$$(6^{+3}_{-2})\times10^{-6}$~\Msun, which is only marginally less
than that of the other lobe. In summary, the scenario depicted by these
results is that of a lobe whose expansion is delayed, but with physical
properties that are basically the same as those characterising the NE
lobe. We conclude that both lobes are powered by the same accretion
event despite the different size, expansion speed, and morphology. Such a
difference can be explained either by an asymmetry in the way the central
engine feeds the two lobes, or by an interaction with the surrounding
material, which may be denser on the SW side (as suggested in Paper~II). On
the basis of the data available to us, it is impossible to discriminate
between these two hypotheses, but further high-resolution monitoring of
the jet might be helpful.

\section{Summary and conclusions}

We used the VLA to image the ionised jet from \S\ at three bands with
subarcsecond resolution in order to complement our previous multi-epoch observations
of the source (Paper~I and~II). The results confirm the appearance of
a SW lobe, delayed with respect to the NE lobe, and allow us to trace
its expansion and establish that the whole jet is fading away. The two
lobes have different sizes, expansion speeds, and structures, hinting at an
asymmetry in either the engine powering them or the surrounding material. We
used the model developed in our previous papers to fit both the morphology
of the lobes and their continuum spectra between 6~GHz and 22.2~GHz. Our
findings confirm a common origin for the two lobes, both powered by the same
accretion outburst studied by Caratti o Garatti et al.~(\cite{cagana}). We
also find that the radio jet is recollimating and recombining.

\begin{acknowledgements}
This study is based on observations made under project 23A-021
of the VLA of NRAO. The National Radio Astronomy Observatory
is a facility of the National Science Foundation operated under cooperative
agreement by Associated Universities, Inc..
A.C.G. acknowledges support from PRIN-MUR 2022 20228JPA3A “The path to
star and planet formation in the JWST era (PATH)” and by INAF-GoG 2022
“NIR-dark Accretion Outbursts in Massive Young stellar objects (NAOMY)”
and Large Grant INAF 2022 “YSOs Outflows, Disks and Accretion: towards
a global framework for the evolution of planet forming systems (YODA)”.
R.F. acknowledges support from the grants Juan de la Cierva
FJC2021-046802-I, PID2020-114461GB-I00 and EX2021-001131- S
funded by MCIN/AEI/10.13039/501100011033 and by “European Union
NextGenerationEU/PRTR”
\end{acknowledgements}

\end{document}